\documentclass{aastex62}

\usepackage{color}
\usepackage{natbib}
\usepackage{graphicx}	

\renewcommand{\baselinestretch}{1.5}
\usepackage{color}
\newcommand{\prot}{$P_{\rm rot}$}

\newcommand{\rvar}{$R_{\rm var}$}
\newcommand{\teff}{$T_{\rm eff}$}
\newcommand{\logg}{$\log g$}
\newcommand{\feh}{[Fe/H]}
\newcommand{\kepler}{\emph{Kepler}}
\newcommand{\s}{\emph{S}}

\newcommand{\figu}[1]{Figure~\ref{#1}}
\usepackage{nicefrac}
\usepackage{multirow}
\shorttitle{solar-type stars}
\shortauthors{Jinghua Zhang et al.}

\begin{document}
\large{}

\title{Solar-type Stars Observed by LAMOST and \kepler{}}

\correspondingauthor{Alexander I. Shapiro}
\email{shapiroa@mps.mpg.de}

\correspondingauthor{Shaolan Bi}
\email{bisl@bnu.edu.cn}

\author[0000-0002-2510-6931]{Jinghua Zhang}
\affiliation{Department of Astronomy, Beijing Normal University, Beijing 100875, People's Republic of China}
\email{zhangjinghua@mail.bnu.edu.cn}

\author{Alexander I. Shapiro}
\affiliation{Max Planck Institute for Solar System Research, G\"{o}ttingen, D-37077, Germany}

\author{Shaolan Bi}
\affiliation{Department of Astronomy, Beijing Normal University, Beijing 100875, People's Republic of China}

\author{Maosheng Xiang}
\affiliation{Max Planck Institute for Astronomy, K\"{o}nigstuhl 17, D-69117, Heidelberg, Germany}

\author[0000-0002-1299-1994]{Timo Reinhold}
\affiliation{Max Planck Institute for Solar System Research, G\"{o}ttingen, D-37077, Germany}

\author[0000-0002-3243-1230]{Krishnamurthy Sowmya}
\affiliation{Max Planck Institute for Solar System Research, G\"{o}ttingen, D-37077, Germany}

\author{Yaguang Li}
\affiliation{Sydney Institute for Astronomy (SIfA), School of Physics, University of Sydney, NSW 2006, Australia}

\author{Tanda Li}
\affiliation{School of Physics and Astronomy, University of Birmingham, Edgbaston, Birmingham, B15 2TT, UK}

\author{Jie Yu}
\affiliation{Max Planck Institute for Solar System Research, G\"{o}ttingen, D-37077, Germany}

\author{Minghao Du}
\affiliation{Department of Astronomy, Beijing Normal University, Beijing 100875, People's Republic of China}

\author{Xianfei Zhang}
\affiliation{Department of Astronomy, Beijing Normal University, Beijing 100875, People's Republic of China}




\begin{abstract}
Obtaining measurements of chromospheric and photometric activity of stars with near-solar fundamental parameters and rotation periods is important for a better understanding of solar-stellar connection. We select a sample of 2603 stars with near-solar fundamental parameters from the the Large Sky Area Multi-Object Fiber Spectroscopic Telescope (LAMOST)-\kepler{} field and use LAMOST spectra to measure their chromospheric activity  and  \kepler{}  light curves to measure their photospheric activity (i.e. the amplitude of the photometric variability). While the rotation periods of 1556 of these stars could not be measured due to the low amplitude of the photometric variability and highly irregular temporal profile of light curves, 254 stars were further identified as having near-solar rotation periods. We show that stars with near-solar rotation periods have chromospheric activities systematically higher than stars with undetected rotation periods. Furthermore, while the solar level of photospheric and chromospheric activity appears to be typical for stars with undetected rotation periods, the Sun appears to be less active than most stars with near-solar rotation periods (both in terms of photospheric and chromospheric activity).
\end{abstract}


\section{Introduction} \label{sec:intro}
The action of a dynamo generates magnetic field in the stellar interior \citep{Charbonneau2010,Paul_book}.
This field emerges in the stellar atmosphere leading to various manifestations of magnetic activity, e.g. brightness and spectroscopic variability, chromospheric and coronal emission. The interest in stellar magnetic activity has been recently brought to a new level by the advent of high-precision transit photometry and subsequent discovery of thousands of exoplanets. This is particularly due to stellar magnetic activity appearing to be a limiting factor in exoplanet detection and characterization, and due to its possible influence on exoplanet atmospheres \citep[see, e.g.][]{Wood2004,Vidotto2019}.

Among the most typical proxies of stellar magnetic activity are chromospheric Ca {\sc ii} \emph{H} and \emph{K} line emission \citep[see, e.g.,][and references therein]{Noyes1984, Baliunas1995, hall2007} and amplitude of the photometric brightness variations \citep{Basri2013,Timo2017}. During the past few decades, the long-term synoptic HK projects at Mount Wilson Observatory \citep[MWO;][]{Wilson1978} and at the Lowell Observatory \citep[][]{hall2007} have been dedicated to measuring the Ca {\sc ii} \emph{H} and \emph{K} emission in late-type stars.
Contemporary to monitoring of the stellar Ca {\sc ii} activity, Lowell and Fairborn Observatories initiated programs for measuring photometric variability of Sun-like stars \citep{radicketal1998, lockwoodetal2007, halletal2009, Radick2018}. Interestingly, it was found that stars with near-solar level of chromospheric activity appear to be much more photometrically variable than the Sun on the decadal timescale. Several explanations of such a puzzle have been proposed, e.g. \cite{witzkeetal2018} suggested that the Sun corresponds to a local minimum of the complex dependence of the amplitude of brightness variations on the activity cycle timescale on fundamental stellar parameters and magnetic activity level.

While the MWO, Lowell, and Fairborn observations formed the backbone of stellar activity studies, they were limited to just a few hundreds of stars. The high-precision photometry from space telescopes, particularly from the \kepler{} mission which observed almost two hundred thousands of stars, allowed to circumvent this limitation. Also, the studies aimed at the comparison of solar and stellar variability enjoyed a breath of fresh air.
In one of the first solar-stellar comparison studies based on the \kepler{} data, \citet{Gilliland2011} suggested that the Sun is photometrically quieter than other presumably main sequence \kepler{} stars with near-solar effective temperatures. Conversely, \citet{Basri2013} found that photometric variability of the Sun is similar to the level of variability displayed by the majority of \kepler{} stars with near-solar effective temperatures. \citet{Salabert2016} identified a sample of 18 solar analogs and found that their photometric variability and chromospheric activities are similar to those of the Sun.

Recently, \citet{Timo2020} combined Gaia and \kepler{} data to identify a sample of \kepler{} stars with effective temperatures between 5500 K and 6000 K. 369 of these stars had rotation periods between
20 and 30 days, while rotation periods of 2529 stars could not be determined from the photometry.
The  photometric variability of the stars with non-detected periods appeared to be very similar to that of the Sun. This is not surprising, since the highly irregular temporal profile of solar brightness variations would make the detection of the solar rotation period from the photometric time-series very difficult \citep[see, e.g.][and references therein]{Aigrain2015,Veronika2020}. Consequently, if the Sun were observed by \kepler{} it would most probably be attributed to a sample of stars with near-solar fundamental parameters but undetected rotation periods. Interestingly, \citet{Timo2020} found that despite having near-solar fundamental parameters and rotation periods, 369 stars with known rotation periods are significantly more variable than the Sun (e.g. the mean variability of the stars with detected periods is almost 5 times larger than solar median variability). Currently, the high variability of these stars and a regular pattern of their light curves remain unexplained. One of the important questions is whether there is any systematic difference between Ca {\sc ii} \emph{H} and \emph{K} emission of the stars with non-detected rotation periods and the Sun on one side and stars with detected near-solar rotation periods on the other side. This question is addressed in this Letter utilizing data from the Large Sky Area Multi-Object Fibre Spectroscopic Telescope (LAMOST) spectroscopic survey \citep{Cui2012, Zhao2012}.

LAMOST has collected millions of stellar spectra with a mean spectral resolution of about 1800 in broad wavelength range of 3700--9100\,\AA{} \citep{Zhao2012}. In particular, it provides vast amounts of Ca {\sc ii} \emph{H} and \emph{K} observational data. Meanwhile, LAMOST spectra allowed accurate determination of stellar fundamental parameters, i.e. effective temperature \teff{}, surface gravity \logg{}, and metallicity [Fe/H]. LAMOST has performed spectroscopic follow-up for targets in the \kepler{} field of view, which was initiated as the LAMOST-\kepler{} project (LK-project) \citep{Decat2015}. By June 2017, this project obtained more than 227,000 low-resolution spectra \citep{Zong2018}, thus providing us with a tool required to answer the question raised above.

In this Letter, we select solar-type stars observed by both LAMOST and \kepler{}, and study the relation between period detectability and chromospheric activity. In Section \ref{sec:data_analysis}, we explain the sample selection and the procedure for measuring the Ca {\sc ii} \emph{H} and \emph{K} emission from the LAMOST spectra. In Section \ref{sec:distribution}, we compare the Ca {\sc ii} \emph{H} and \emph{K} emission of the Sun and stars with known and unknown rotation periods. We summarise our results in Section \ref{sec:conclusion}.

\section{Data and methods} \label{sec:data_analysis}
\subsection{Stellar samples}\label{sec:data}
The stars analyzed in this study have been selected from the sixth Data Release (DR6) of the LAMOST survey\footnote{\url{http://dr6.lamost.org/}}. Here, we define solar-type stars as stars with \teff{}, \logg{} and \feh{} in the ranges 5500--6000 K, 4.14--4.74 and $-0.2$--0.2, respectively. All these parameters are taken from DR6 of the LAMOST survey which is based on the LAMOST stellar parameter pipeline \citep[LASP;][]{Zhao2012, Luo2015}. The solar values were taken to be: \teff{} = 5777 K and \logg{} = 4.44. To place a lower limit on the quality of the spectroscopic observations, we only considered stars with the signal-to-noise ratios (S/Ns) at the blue end of the spectra higher than 30. With these constraints, we collected 341,557 spectra for 272,854 solar-type stars, which we denote as LAMOST sample. Among them, there are 6626 stars which have been also observed by \kepler{} mission \citep{Zong2018}, which we denote as the L-K sample.

We cross-matched the selected 6626 stars with the catalog of \citet{McQuillan2014}. This is a catalog of \kepler{} stars containing 34,030 stars with detected rotation periods and 99,000 stars with non-detected rotation periods.  Following \citet{Timo2020} we concentrated on stars with rotation periods in the range 20--30 days (hereafter, solar-type stars) and stars with non-detected rotation period (hereafter, non-periodic sample). Furthermore, we have also selected stars with periods in the range 10--20 days (hereafter, short-periodic sample).  Such a classification results in 254 solar-type stars and 793 short-periodic stars. 1556 stars were deemed as non-periodic. These stars can be considered as pseudo solar-type stars since their rotation periods are unknown. As discussed in Section \ref{sec:intro} the Sun would most probably be allocated to the non-periodic sample of pseudo solar-type stars. In the Appendix we also consider stars with rotation periods shorter than 10 days to better illustrate the effect of the rotation period on photometric variability and on Ca {\sc ii} \emph{H} and \emph{K} emission.

\subsection{Chromospheric activity}\label{sec:analysis}
Using the LAMOST spectra, we measured the magnetic activity proxy \s{}-index as:
  \begin{equation}\label{1}
    S_{\rm LAMOST} = \alpha\cdot\frac{H + K}{R + V} ,
  \end{equation}
where \emph{H} and \emph{K} are the integrated fluxes in the cores of Ca {\sc ii} \emph{H} and \emph{K} lines, respectively. The integration is performed using a triangle function with a full width at half-maximum (FWHM) of 1.09\,\AA{} centered at 3968\,\AA{} and 3934\,\AA{}, respectively. The parameters \emph{R} and \emph{V} are the integrated fluxes in the nearby pseudo-continuum. The integration in pseudo-continuum is performed using a rectangular function with 20\,\AA{} width centered at 4001\,\AA{} and 3901\,\AA{}, respectively. Following \citet{Karoff2016} we put calibration factor $\alpha$ to 14.4. \citet{Karoff2016} argued that such a choice of the calibration factor leads to a distribution of LAMOST \s{}-index values being consistent with that derived from \cite{Isaacson2010} around $\s{}=0.2$. For stars with multiple observations, the \s{}-indexes were determined by using the weighted mean values of these multiple spectra with the weights being the S/Ns of the spectra. We refer to \citet{Zhang2020} for a detailed discussion of \s{}-index measurements.

One of the main limitations of the current study is that \kepler{}  and LAMOST  observations are not performed at the same moment in time.  While \kepler{} data considered in this study were obtained from June 2009 till May 2013, most of the LAMOST spectra were taken from June 2012 to June 2017. Consequently, 92\% of LAMOST spectra used in this study have been taken outside of the period of \kepler{} observations. Nevertheless, we do not expect \s{}-index  to change significantly between the periods of  \kepler{} and LAMOST observations.
Indeed, the amplitude of the \s{}-index  variability on the rotation and activity cycle timescales is proportional to the time-averaged values of the \s{}-index \citep[see, e.g.,][and references therein]{Ricky_PhD, Radick2018}. Consequently, the changes of \s{}-index values of stars in the solar-type and non-periodic samples are expected to be similar to those of the Sun (i.e. about 10\% from the mean value).  Furthermore, spectra of 577 stars in our samples have been recorded by LAMOST more than once with the intervals among observations often reaching a couple of years. Comparison of these spectra did not reveal any significant changes of the corresponding \s{}-index values with time (e.g. standard deviation among  \s{}-index values for the majority of the non-periodic stars was below 0.01--0.015).

\section{Results} \label{sec:distribution}
\subsection{\s{}-index distributions} \label{sec:solar-type distribution}
\begin{figure}
\figurenum{1}
\plotone{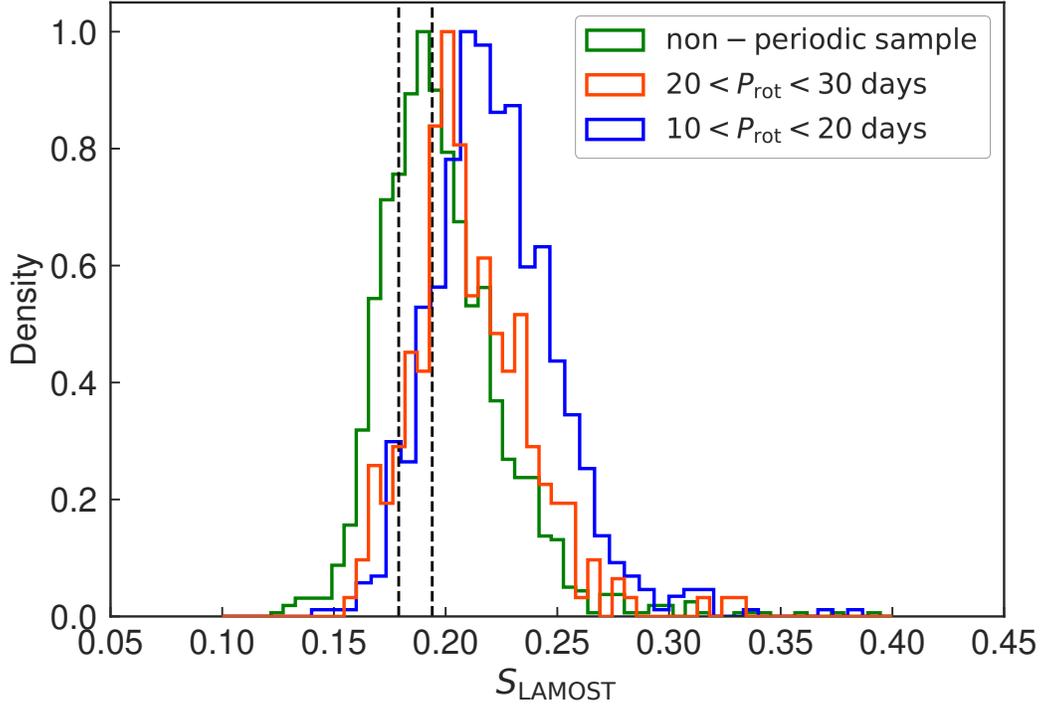}
\caption{The distributions of the \s{}-index values for the short-periodic sample (blue), solar-type sample (red), and non-periodic sample (green). The dashed vertical lines indicate the minimum and maximum values of the solar \s{}-index as it would be measured by LAMOST during activity cycles 15--24 (see Sect.~\ref{sec:solar distribution} for the detailed explanation). \label{fig:hist_mc14}}
\end{figure}

\figu{fig:hist_mc14} shows distributions of \s{}-index values for three stellar samples introduced in Section~\ref{sec:data}.
Not surprisingly the short-periodic sample appears to be on an average more active than the sample of solar-type stars. Interestingly, \figu{fig:hist_mc14} indicates that non-periodic stars are on an average less active than solar-type stars. Furthermore, the difference between distributions of solar-type and non-periodic stars is similar to the difference between distributions of short-periodic and solar-type stars.

To quantify the difference between the distributions we computed the Kolmogorov-Smirnov (K-S) statistic \citep{hodges1958}. The K-S statistic are measured by the maximum diagonal distance between the empirical cumulative distribution functions of the two samples. The statistic is 0.21 for distributions between short-periodic and solar-type sample while it is 0.27 for distributions between solar-type and non-periodic sample. The K-S statistic values and the two low \emph{p}-values ($< 10^{-6}$) indicate distributions of both sample sets are different, and distributions between the solar-type and non-periodic sample are much more different than those between short-periodic and solar-type sample.

\begin{figure}
\figurenum{2}
\plotone{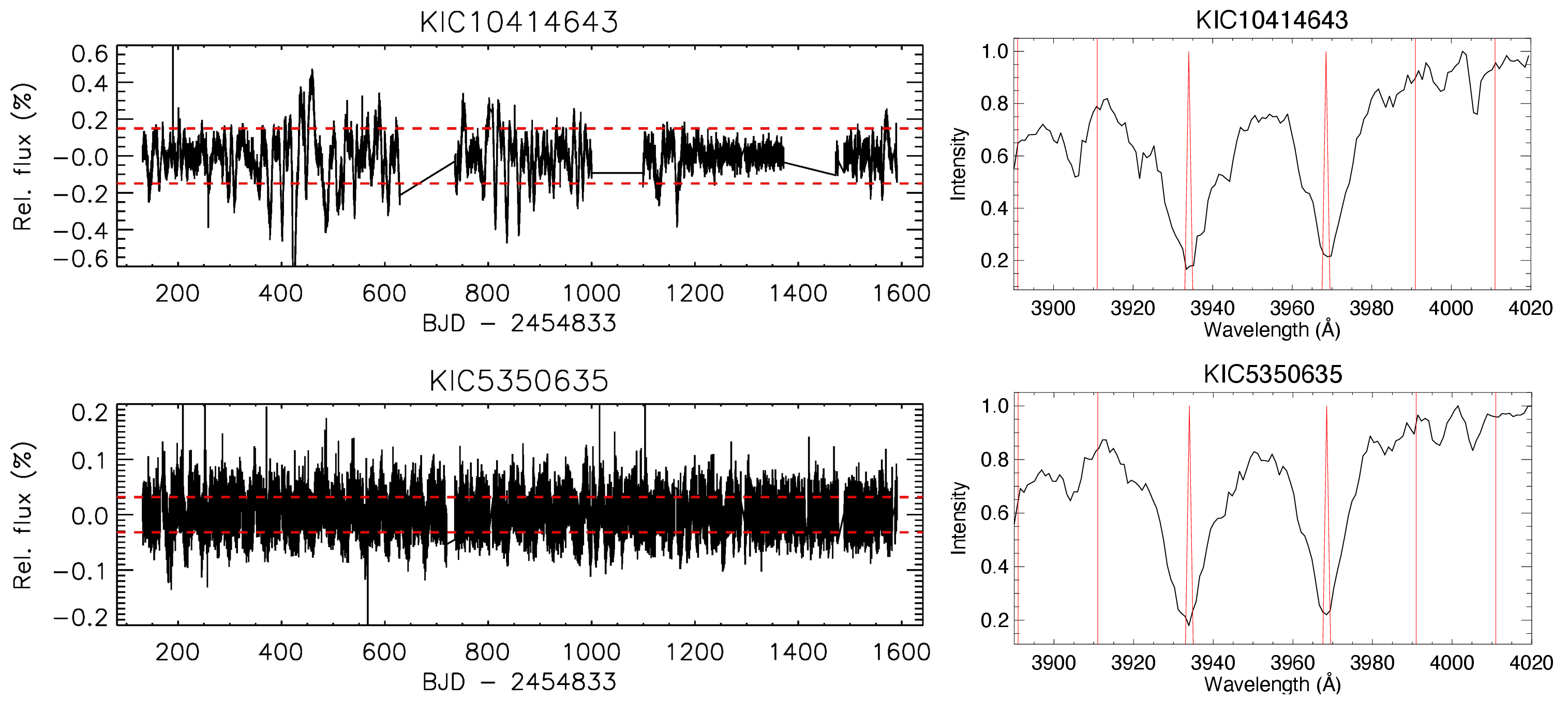}
\caption{The \kepler{} light curves (left panels) and LAMOST Ca {\sc ii} \emph{H} \& \emph{K} profiles (right panels) for solar-type  star KIC\,10414643 (top panels) and non-periodic star KIC\,5350635 (bottom panels), respectively. The horizontal red dashed lines in the two left panels indicate the variability range \rvar{}. The red triangles in the two right panels indicate the measurement bandpasses at the cores of Ca {\sc ii} \emph{H} \& \emph{K} lines, the red rectangles indicate the measurement bandpasses of pseudo-continuum. \label{fig:light_curve}}
\end{figure}

\figu{fig:hist_mc14} indicates that though the \s{}-index distributions of solar-type and non-periodic stars are different, there is a substantial overlap between them. This implies that stars with detectable and not-detectable rotation periods and, consequently, very different light curves can still have the same levels of the chromospheric activity. The example of two such stars, KIC\,10414643 and KIC\,5350635, is given in Figure~\ref{fig:light_curve} and their main parameters are summarised in Table~\ref{tab:sample}. The photometric variation \rvar{} is taken from \citet{Timo2020}. It is calculated by defining the difference between the 5th and 95th percentile of the sorted differential flux in each \kepler{} quarter and then taking the median among all quarters value.

\figu{fig:light_curve} shows that while both stars have very similar Ca {\sc ii} \emph{H} and \emph{K} profiles and basically the same values of the \s{}-index, their light curves are pretty different. The light curve of KIC\,10414643, on the one hand, is highly regular and its rotation period can be easily determined. On the other hand, the amplitude of variability of KIC\,5350635 is roughly five times smaller and no clear periodic signal can be seen behind the noise.

\begin{deluxetable}{ccccccccccccc}
\tablecaption{Parameters of  KIC\,10414643 and  KIC\,5350635. \label{tab:sample}}
\tablehead{
KIC & \teff{} & \logg{} & [Fe/H] & \rvar{}  & \s{}-index  & \prot{} }
\startdata
10414643&$5692.39\pm30.79$&$4.50\pm0.05$&$0.02\pm0.03$&$0.2991$&$0.2018$&$22.21$\\
5350635&$5829.83\pm23.08$&$4.46\pm0.04$&$-0.11\pm0.02$&$0.0638$&$0.2019$& \\
\enddata
\tablecomments{}{The values of effective temperature, surface gravity, and metallicity are take from DR6 of the LAMOST survey. The values of photometric variations are taken from \citet{Timo2020}. The value of rotation period is taken from \citet{McQuillan2014}.}
\end{deluxetable}

One possible explanation of such behaviour is that chromospheric activity is mainly given by the overall coverage of a star by the magnetic features \citep[see, e.g.,][]{Shapiro2014_stars}. At the same time, the photometric variability on stellar rotation timescale strongly depends on the surface distribution of magnetic features, their sizes, as well as evolution \citep[see, e.g.,][]{Shapiro2020a}. In particular, recent studies have been able to reveal the temporal evolution of starspots \citep[see, e.g.][]{Namekata2020}
and also determine their sizes \citep[see, e.g.][who showed that spots of HAT-P-11 and of the Sun have similar sizes]{Morris2017}.

One interesting effect capable of a strong increase of the photometric variability without a direct influence on the \s{}-index values and, thus, explaining the difference between periodic and non-periodic stars is nesting in the distribution of magnetic features \citep[i.e. the tendency of magnetic features to emerge within certain ``nests'' of activity, see, e.g.][]{nests}. In contrast to the spatially random distribution of emergences, nesting would lead to a non-axisymmetric distribution of spots \citep[see, e.g., ][]{Emre2018}  and, consequently, regular light curves with large amplitudes of the rotational brightness variability. The effect of nesting on the photometric variability will be addressed in the forthcoming study. \footnote{E.Isik(2020, private communication).}

Another contributing factor might be the stellar inclination, i.e. the angle between the direction to the observer and stellar rotation axis. While photometric variability and period detectability strongly depends on the inclination of a star \citep{Nina2020}, chromospheric activity shows a much weaker dependence \citep{Shapiro2014_stars}. In particular, if a star is observed at a relatively low inclination (i.e. pole-on) its rotational variability would be significantly reduced and a star will be classified as non-periodic despite a large chromospheric activity. Finally, we can not fully exclude a possible change of the \s{}-index between periods of \kepler{} and LAMOST observations (although, see the discussion in Sect.~\ref{sec:analysis}). We emphasize here the need for the future contemporaneous spectroscopic and photometric observations for a large sample of stars.

\subsection{Relation between \s{}-index and photometric variability} \label{sec:S_Rvar}
\begin{figure}
\figurenum{3}
\plotone{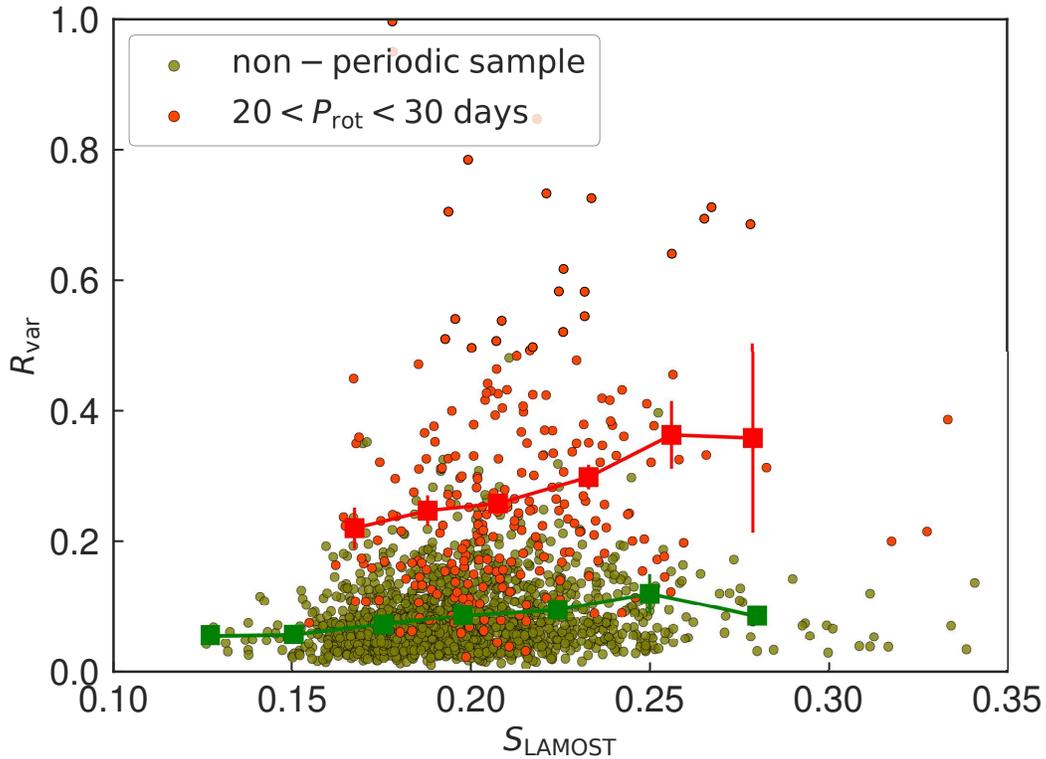}
\caption{Dependence of photometric variability, \rvar{}, on \s{}-index for the  solar-type sample (red) and non-periodic sample (green). The red and green square symbols connected with lines represent the averaged \rvar{} and \s{}-index values in 7 bins (see Section~\ref{sec:S_Rvar} for more details) for the solar-type sample and non-periodic sample, respectively. The vertical line segments indicate standard deviation of \rvar{} values within the bin. \label{fig:S_lamost_Rvar}}
\end{figure}

In \figu{fig:S_lamost_Rvar} we plot the dependences of photometric variability on \s{}-index for the solar-type and non-periodic samples. For the solar-type sample, the \rvar{} significantly increases with \s{}-index. We binned the \s{}-index values into 7 equidistant segments within the range 0.12--0.30 for the solar-type sample and within the range 0.1--0.3 for the non-periodic sample. The averaged \rvar{} values in each bins were then calculated. The binned values show that for both samples photometric variability somehow increases with the \s{}-index. The increase is, however, not particularly strong and is to a large extent hidden by the large spread of photometric variabilities the stars with the same \s{}-index can have.

In agreement with \citet{Timo2020}, \figu{fig:S_lamost_Rvar} shows that photometric variability of solar-type stars is significantly larger than that of non-periodic stars. The stars in the non-periodic sample exhibit variabilities similar to that of the Sun \citep[for which median \rvar{} over the last 140 years was 0.07\% and maximum \rvar{} was about 0.2\%, see][for detailed discussion]{Timo2020}. At the same time \figu{fig:S_lamost_Rvar} demonstrates that although distributions of \s{}-index values for the periodic and non-periodic stars are different (see Section.~\ref{sec:solar-type distribution}), the \s{}-index is not the main factor which defines the morphology of the light curves (regular vs. non-regular) and their amplitude. We note that the \s{}-index mainly depends on the total coverage of stellar surface by magnetic features, while the photometric variability additionally depends on the degree of axisymmetry of this distribution (see Section.~\ref{sec:solar-type distribution}).
Consequently, our result hints that the surface distribution of magnetic features and its degree of axisymmetry plays an important role in defining whether a star is solar-type or non-periodic stars.

\kepler{} data shows that photometric variability increases with the
stellar rotation rate until periods of about 12 days and saturates for faster rotators \citep[see, e.g., Fig.~15 in][]{Notsu2019}. Interestingly, the saturation appears to be less pronounced for the \s{}-index values \citep[see Fig.~6 in][]{Zhang2020}. We further illustrate this point in the Appendix where we repeat Figs.~\ref{fig:hist_mc14}~and~\ref{fig:S_lamost_Rvar} but for samples of stars with rotational periods $P_{\rm rot}<8$~days,  $8<P_{\rm rot}<12$~days, $12<P_{\rm rot}<20$~days, and $20<P_{\rm rot}<30$~days (see Figs.~\ref{fig:hist_appendix}~and~\ref{fig:S_lamost_Rvar_v}). One can see that \s{}-index values and photometric variabilities for the last three samples are clearly different, increasing from slower to faster rotators (see red, blue, and maroon in Figs.~\ref{fig:hist_appendix}~and~\ref{fig:S_lamost_Rvar_v}). For the \s{}-index the same trend is also valid for stars with $P_{\rm rot}<8$~days - they appear to be more active than stars in other samples (see yellow in  Fig~\ref{fig:hist_mc14}). At the same time there seems to be a saturation in the photometric variabilities since  their average values for stars with $P_{\rm rot}<8$ days and for stars with  $8<P_{\rm rot}<12$ days are very similar (see maroon and yellow in Fig.~\ref{fig:S_lamost_Rvar_v}). This is consistent with the
results of \cite{Notsu2019} and \cite{Zhang2020}.

\subsection{Effect of spectral resolution on the \s{}-index and placing the Sun among solar-type stars} \label{sec:solar distribution}
The main difficulty in finding the solar \s{}-index as it would be measured by LAMOST is the relatively low spectral resolution of LAMOST. Although the mean resolution power of LAMOST is about 1800, the  resolution strongly depends on wavelength and is expected to be about 1000 around Ca {\sc ii} \emph{H} and \emph{K}  lines \citep{Xiang2015}. Furthermore, \citet{Xiang2015} reported that the resolution power of each LAMOST fiber varies considerably, with amplitudes amounting to 1\,\AA{}. Potentially such a spread of resolutions might affect not only placing the Sun on the LAMOST \s{}-index scale but also the distributions plotted in \figu{fig:hist_mc14}. Indeed, one might expect that stars with near-solar activity levels observed at lower spectral resolution will have higher \s{}-index values than stars observed at higher resolution (since the lower the spectral resolution is the stronger the line cores are mixed with wings and, consequently, the larger are the \emph{H} and \emph{K} fluxes , see Equation \ref{1}).

\begin{figure}
\figurenum{4}
\plotone{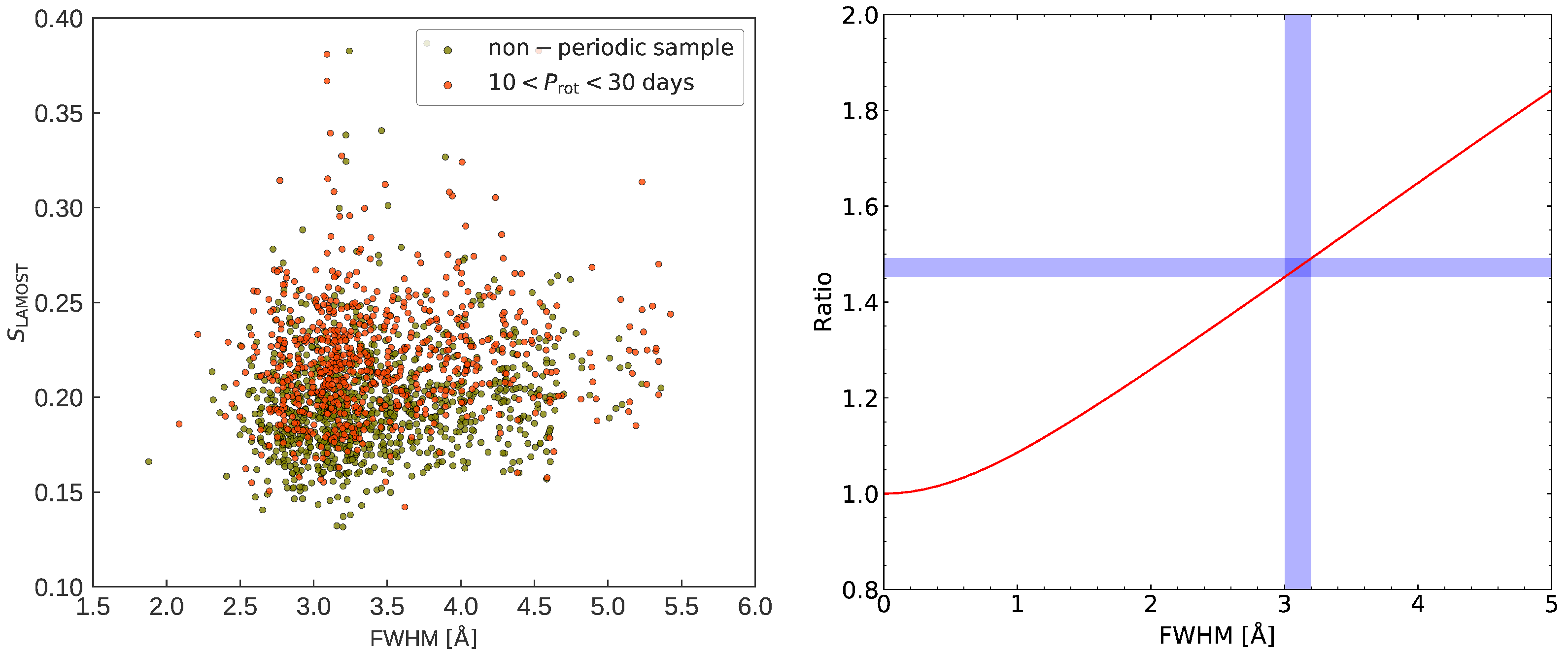}
\caption{Left panel: the \s{}-index values determined from the LAMOST spectra vs. spectral resolution power of fibres used to obtain these spectra. Right panel: modelled ratio between solar \s{}-index calculated from the FTS solar spectrum convolved with Gaussian kernel of corresponding FWHM and \s{}-index calculated with FTS spectrum without any convolution.
The vertical shaded area in the right panel indicates the FWHM in range 3.0--3.2 where most stars from our samples were observed. The horizontal shaded area indicates the corresponding ratio in range 1.45--1.49. \label{fig:fwhm}}
\end{figure}

To clarify how strong the impact of the differences in the resolution power between different LAMOST fibres on the distribution of \s{}-index values is, we collected the observational resolution curves at the blue end (3700--5900\,\AA{}) of the individual spectra which were obtained utilizing the LAMOST arc lamp and sky emission lines.\footnote{M.-S., Xiang(2020, private communications).} In the left panel of Figure \ref{fig:fwhm} we show the dependence of the \s{}-index on the LAMOST observational resolution power, i.e. FWHM for stars from \figu{fig:hist_mc14}. Note that the observational resolution curves are available for $\sim$50\% of stars in \figu{fig:hist_mc14}.
One can see that  the dependence (if any) of the \s{}-index on the LAMOST spectral resolution is rather weak  compared to the differences of \s{}-index values between the periodic and non-periodic sample. Consequently, we do not expect that any of our results might be affected by the LAMOST stars being observed at slightly different spectral resolutions.

It is important to place the Sun and solar-type stars on the same \s{}-index scale for comparing their chromospheric activities. This, however, is hindered by the fact that LAMOST survey has no record of spectra of solar light reflected from minor bodies or from inactive satellites. Therefore, we take the following indirect approach to obtain the \s{}-index of the Sun on LAMOST scale. We use the high resolution (better than 350 000) solar flux spectrum from Hamburg atlas\footnote{\url{ftp://ftp.hs.uni-hamburg.de/pub/outgoing/FTS-Atlas/}} \citep[see, for e.g.][]{2016A&A...590A.118D} created using the data from the Fourier Transform Spectrometer (FTS) at the McMath-Pierce solar telescope at the Kitt Peak National Observatory. We degrade this FTS spectrum by convolving it with Gaussian kernels of varying FWHM in the range 0--5\,\AA{}, which covers the range of spectral resolutions achieved from LAMOST fibres. Then we compute the \s{}-index for the high resolution (undegraded) and degraded spectra. In the right panel of \figu{fig:fwhm} we show the ratio of \s{}-index measured for the degraded spectrum to that measured for the original high resolution FTS spectrum as a function of the resolution. Due to the increased \emph{H} and \emph{K} fluxes resulting from the lower spectral resolution, the ratio increases with decreasing resolution. In the left panel of \figu{fig:fwhm}, it appears that the majority of LAMOST stars considered for this plot were observed with spectral resolution between 3.0 and 3.2\,\AA{}. For this range of resolution, the ratio varies only slightly i.e. between 1.45--1.49 as indicated by the shaded horizontal bar in the right panel of \figu{fig:fwhm}, reinforcing that \s{}-index has a weaker dependence on the variations in LAMOST resolution. However, we note that \s{}-index measured with a low resolution spectra like LAMOST would be on an average larger by 47\%\ than that measured with the high resolution spectrum.

\cite{Egeland2017} has accurately placed the solar \s{}-index value on the MWO \s{}-index scale. They found that minimum and maximum values of the solar \s{}-index during cycles 15--24 were 0.162 and 0.177, respectively. We note that these values correspond to $\alpha=19.2$ instead of  $\alpha=14.4$ we adopted here following \citet{Karoff2016}. Thus we first corrected \cite{Egeland2017} values for the difference in calibration factors and then applied factor 1.47 (see right panel of \figu{fig:fwhm}) to correct for LAMOST spectral resolution. As a result we find that solar \s{}-index on the LAMOST scale varied  between 0.179 and 0.194 during cycles 15--24. These values are designated by the dashed vertical lines in Figure \ref{fig:hist_mc14}. Compared to the activity level of the Sun, the peak level of activity of the non-periodic stars is near the range of the Sun, while the majority of the periodic stars have activity levels higher than that of the Sun. Consequently, the analysis of the Ca {\sc ii} \emph{H} and \emph{K} data reinforces the conclusion of \cite{Timo2020} (drawn from the analysis of the \kepler{} light curves) that the Sun is a typical star of the non-periodic sample.

\section{conclusions} \label{sec:conclusion}
We derived the chromospheric activity indexes of 2603 stars in LAMOST-\kepler{} project. These stars were classified into three different samples. The solar-type sample includes stars with known rotation periods in range of 20--30 days, the short-periodic sample includes stars with known rotation periods in range of 10--20 days, and the non-periodic sample includes stars with unknown rotation periods. We investigated the \s{}-index distributions of these samples. We studied the dependence of the photometric variation on the chromospheric activity level for the solar-type sample and non-periodic sample. By convolving the high-resolution solar spectrum to the LAMOST resolution, we could place the Sun on the LAMOST \s{}-index scale.

We showed that the solar \s{}-index values are typical for the non-periodic \kepler{} stars. In contrast, the stars in the solar-type sample are systematically more active than the non-periodic stars in both chrmospheric activity levels and amplitudes of photometric variation. At the same time we found that non-periodic and solar-type stars can have the same values of the \s{}-index. Consequently, the \s{}-index, which mainly determines the total coverage of stellar surface by magnetic features, is not the main factor determining stellar photometric variability. We suggest that the  surface distribution of magnetic features and, in particular, the degree of its axisymmetry plays at least as important role in defining stellar photometric variability as the total coverage of a star by magnetic features.

\vspace{7mm} \noindent {\bf Acknowledgments}
This work is supported by the Joint Research Fund in Astronomy (U1631236) under cooperative agreement between the National Natural Science Foundation of China (NSFC) and Chinese Academy of Sciences (CAS), and grants 11903044 from the NSFC.  Shapiro A.I. and Reinhold T. have been funded by the European Research Council (ERC) under the European Union's Horizon 2020 research and innovation programme (grant agreement No.  715947). Xiang, M.-S. acknowledges support from NSFC Grant No.11703035. Sowmya, K. received funding from the European Union's Horizon 2020 research and innovation
programme under the Marie Sk{\l}odowska-Curie grant agreement No. 797715. T.L. acknowledge the funding from the European Research Council (ERC) under the European Union's Horizon 2020 research and innovation programme (CartographY GA. 804752). Zhang, X.-F. acknowledges support from NSFC Grant No.11703001.
We acknowledge the entire \kepler{} team and everyone involved in the \kepler{} mission. Funding for the \kepler{} Mission is provided by NASA’s Science Mission Directorate.
Guoshoujing Telescope (the Large Sky Area Multi-Object Fiber Spectroscopic Telescope, LAMOST) is a National Major Scientific Project built by the Chinese Academy of Sciences.
Funding for the project has been provided by the National Development and Reform Commission. LAMOST is operated and managed by the National Astronomical Observatories, Chinese Academy of Sciences.

\appendix

\begin{figure}
\figurenum{A1}
\plotone{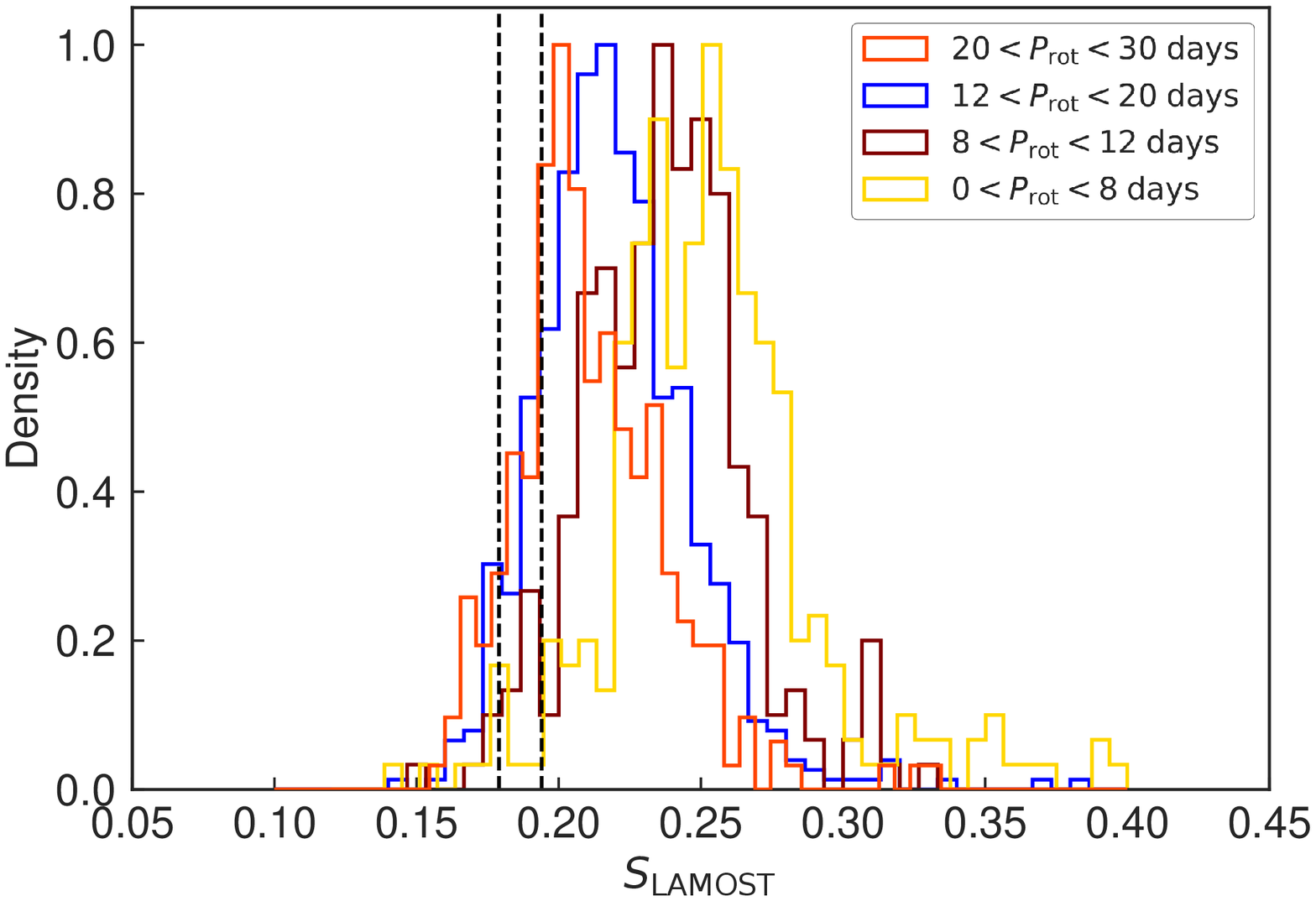}
\caption{The same as Fig.~\ref{fig:hist_mc14} from the main text but for different samples of stars (see the figure legend). \label{fig:hist_appendix}}
\end{figure}

\begin{figure}
\figurenum{A2}
\plotone{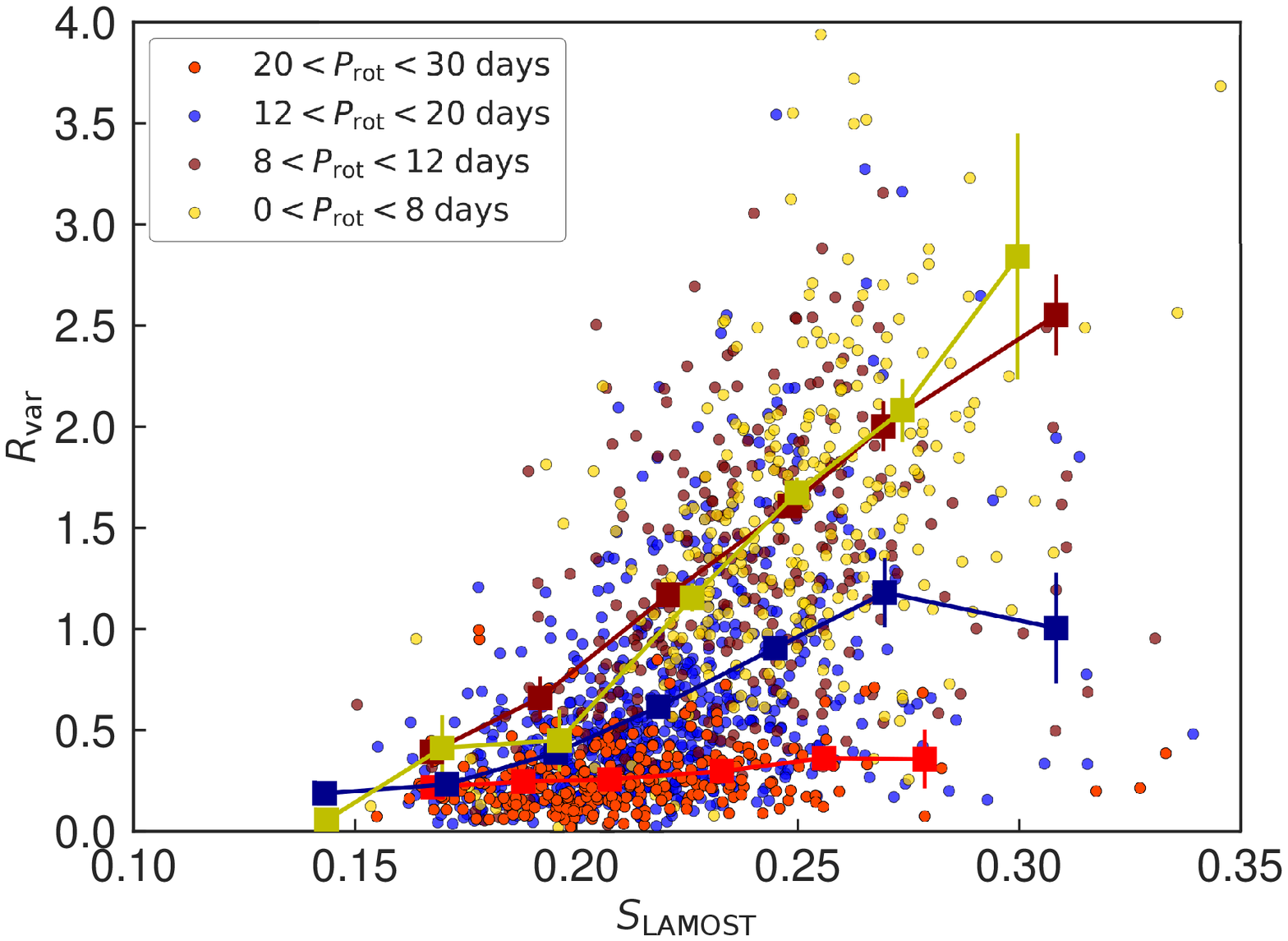}
\caption{The same as Fig.~\ref{fig:S_lamost_Rvar} from the main text but for different samples of stars (the samples shown here are the same as in Fig.~\ref{fig:hist_appendix}, see the figure legend). \label{fig:S_lamost_Rvar_v}}
\end{figure}

Distributions of \s{}-index values (\figu{fig:hist_appendix}) and the dependence of photometric variability \rvar{} on \s{}-index (\figu{fig:S_lamost_Rvar_v}) for periodic stars grouped according to their rotation periods.

\clearpage



\bibliographystyle{plainnat}
\bibliography{ref} 

\label{lastpage}

\end{document}